\documentstyle[prl,floats,eqsecnum,aps,epsfig,bm]{revtex}

\begin{document}
\def\today{\space\number\day\space\ifcase\month\or January\or February\or
  March\or April\or May\or June\or July\or August\or September\or October\or
  November\or December\fi\space\number\year}

 \twocolumn[\hsize\textwidth\columnwidth\hsize\csname@twocolumnfalse\endcsname

\title{Measurement of the rate of $\nu_e+d\rightarrow p+p+e^-$ 
interactions produced by $^{\bm 8}$B solar neutrinos at the Sudbury
Neutrino Observatory}  

\author{
Q.R.~Ahmad$^{15}$,
R.C.~Allen$^{11}$,
T.C.~Andersen$^{12}$,
J.D.~Anglin$^7$,
G.~B\"uhler$^{11}$,
J.C.~Barton$^{13}$\cite{Birkbeck}, 
E.W.~Beier$^{14}$,
M.~Bercovitch$^7$,
J.~Bigu$^4$,
S.~Biller$^{13}$, 
R.A.~Black$^{13}$, 
I.~Blevis$^3$,
R.J.~Boardman$^{13}$, 
J.~Boger$^2$,
E.~Bonvin$^9$, 
M.G.~Boulay$^9$, 
M.G.~Bowler$^{13}$, 
T.J.~Bowles$^6$,
S.J.~Brice$^{6,13}$,
M.C.~Browne$^{15}$,
T.V.~Bullard$^{15}$,
T.H.~Burritt$^{15,6}$,
K.~Cameron$^{12}$,
J.~Cameron$^{13}$, 
Y.D.~Chan$^5$,
M.~Chen$^9$,
H.H.~Chen$^{11}$\cite{dec},                 
X.~Chen$^{5,13}$,
M.C.~Chon$^{12}$,
B.T.~Cleveland$^{13}$,
E.T.H.~Clifford$^{9,1}$,
J.H.M.~Cowan$^4$,
D.F.~Cowen$^{14}$,
G.A.~Cox$^{15}$,
Y.~Dai$^9$,  
X.~Dai$^{13}$, 
F.~Dalnoki-Veress$^3$,
W.F.~Davidson$^7$,
P.J.~Doe$^{15,11,6}$,
G.~Doucas$^{13}$,
M.R.~Dragowsky$^{6,5}$,
C.A.~Duba$^{15}$,
F.A.~Duncan$^9$,
J.~Dunmore$^{13}$, 
E.D.~Earle$^{9,1}$, 
S.R.~Elliott$^{15,6}$,
H.C.~Evans$^9$,
G.T.~Ewan$^9$, 
J.~Farine$^3$,
H.~Fergani$^{13}$,
A.P.~Ferraris$^{13}$,
R.J.~Ford$^9$, 
M.M.~Fowler$^6$,
K.~Frame$^{13}$,
E.D.~Frank$^{14}$,
W.~Frati$^{14}$,
J.V.~Germani$^{15,6}$,
S.~Gil$^{10}$,
A.~Goldschmidt$^6$,
D.R.~Grant$^3$,
R.L.~Hahn$^2$,
A.L.~Hallin$^9$, 
E.D.~Hallman$^4$,
A.~Hamer$^{6,9}$,
A.A.~Hamian$^{15}$,
R.U.~Haq$^4$,
C.K.~Hargrove$^3$,
P.J.~Harvey$^9$, 
R.~Hazama$^{15}$,
R.~Heaton$^9$,
K.M.~Heeger$^{15}$,
W.J.~Heintzelman$^{14}$,
J.~Heise$^{10}$,
R.L.~Helmer$^{10}$\cite{TRIUMF},
J.D.~Hepburn$^{9,1}$,
H.~Heron$^{13}$, 
J.~Hewett$^4$,
A.~Hime$^6$,
M.~Howe$^{15}$,
J.G.~Hykawy$^4$,
M.C.P.~Isaac$^5$,
P.~Jagam$^{12}$,
N.A.~Jelley$^{13}$, 
C.~Jillings$^9$, 
G.~Jonkmans$^{4,1}$,
J.~Karn$^{12}$,
P.T.~Keener$^{14}$,
K.~Kirch$^6$,
J.R.~Klein$^{14}$,
A.B.~Knox$^{13}$, 
R.J.~Komar$^{10,9}$,
R.~Kouzes$^8$,
T.~Kutter$^{10}$,
C.C.M.~Kyba$^{14}$,
J.~Law$^{12}$,
I.T.~Lawson$^{12}$,
M.~Lay$^{13}$,
H.W.~Lee$^9$, 
K.T.~Lesko$^5$,
J.R.~Leslie$^9$,  
I.~Levine$^3$,
W.~Locke$^{13}$, 
M.M.~Lowry$^8$,
S.~Luoma$^4$,
J.~Lyon$^{13}$,
S.~Majerus$^{13}$,
H.B.~Mak$^9$, 
A.D.~Marino$^5$,
N.~McCauley$^{13}$,
A.B.~McDonald$^{9,8}$, 
D.S.~McDonald$^{14}$,
K.~McFarlane$^3$,
G.~McGregor$^{13}$, 
W.~McLatchie$^9$, 
R.~Meijer Drees$^{15}$,
H.~Mes$^3$,
C.~Mifflin$^3$,
G.G.~Miller$^6$,
G.~Milton$^1$,
B.A.~Moffat$^9$,
M.~Moorhead$^{13,5}$, 
C.W.~Nally$^{10}$,
M.S.~Neubauer$^{14}$,
F.M.~Newcomer$^{14}$,
H.S.~Ng$^{10}$,
A.J.~Noble$^3$\cite{TRIUMF}, 	
E.B.~Norman$^5$,
V.M.~Novikov$^3$,
M.~O'Neill$^3$,
C.E.~Okada$^5$,
R.W.~Ollerhead$^{12}$,
M.~Omori$^{13}$, 
J.L.~Orrell$^{15}$,
S.M.~Oser$^{14}$,
A.W.P.~Poon$^{5,6,10,15}$,
T.J.~Radcliffe$^9$, 
A.~Roberge$^4$,
B.C.~Robertson$^9$, 
R.G.H.~Robertson$^{15,6}$,
J.K.~Rowley$^2$,
V.L.~Rusu$^{14}$,
E.~Saettler$^4$,
K.K.~Schaffer$^{15}$,
A.~Schuelke$^5$,
M.H.~Schwendener$^4$,
H.~Seifert$^{4,6,15}$,
M.~Shatkay$^3$,
J.J.~Simpson$^{12}$,
D.~Sinclair$^3$,
P.~Skensved$^9$,
A.R.~Smith$^5$,
M.W.E.~Smith$^{15}$,
N.~Starinsky$^3$,
T.D.~Steiger$^{15}$,
R.G.~Stokstad$^5$,
R.S.~Storey$^7$\cite{dec},         
B.~Sur$^{1,9}$,
R.~Tafirout$^4$,
N.~Tagg$^{12}$,
N.W.~Tanner$^{13}$, 
R.K.~Taplin$^{13}$,
M.~Thorman$^{13}$, 
P.~Thornewell$^{6,13,15}$
P.T.~Trent$^{13}$\cite{Birkbeck},
Y.I.~Tserkovnyak$^{10}$,
R.~Van\ ~ Berg$^{14}$,
R.G.~Van de Water$^{14,6}$,  
C.J.~Virtue$^4$,
C.E.~Waltham$^{10}$,
J.-X.~Wang$^{12}$,
D.L.~Wark$^{13,6}$\cite{RAL_Sussex}, 
N.~West$^{13}$,
J.B.~Wilhelmy$^6$,
J.F.~Wilkerson$^{15,6}$,
J.~Wilson$^{13}$,
P.~Wittich$^{14}$,
J.M.~Wouters$^6$,
M.~Yeh$^2$\\
(The SNO Collaboration) 
}
\address{
$^1$ Atomic Energy of Canada Limited, Chalk River Laboratories, Chalk River,
Ontario K0J 1J0 \\
$^2$Chemistry Department, Brookhaven National Laboratory,  Upton, NY
11973-5000 \\
$^3$Carleton University, Ottawa, Ontario K1S 5B6 Canada\\
$^4$Department of Physics and Astronomy, Laurentian University,
Sudbury, Ontario P3E 2C6 Canada\\
$^5$Institute for Nuclear and Particle Astrophysics and Nuclear Science 
Division, Lawrence Berkeley National Laboratory, 
Berkeley, CA 94720 \\ 
$^6$Los Alamos National Laboratory, Los Alamos, NM 87545 \\
$^7$National Research Council of Canada, Ottawa, Ontario K1A 0R6
Canada\\
$^8$Department of Physics, Princeton University, Princeton, NJ 08544\\
$^9$Department of Physics, Queen's University, Kingston, Ontario K7L
3N6 Canada\\
$^{10}$Department of Physics and Astronomy, University of British Columbia,
Vancouver, BC V6T 1Z1 Canada\\
$^{11}$Department of Physics, University of California, Irvine, CA 92717\\
$^{12}$Physics Department, University of Guelph,  Guelph, Ontario N1G 2W1
Canada\\ 
$^{13}$Nuclear and Astrophysics Laboratory, University of Oxford, Keble
Road, Oxford, OX1 3RH, UK\\
$^{14}$Department of Physics and Astronomy, University of Pennsylvania,
Philadelphia, PA 19104-6396, \\
$^{15}$Center for Experimental Nuclear Physics and Astrophysics, and
Department of Physics, University of Washington,
Seattle, WA 98195 
}

\date{\today}
\maketitle
\begin{abstract}
 Solar neutrinos from the decay of $^8$B have been detected
 at the Sudbury Neutrino Observatory (SNO) via the charged
 current (CC) reaction on deuterium and by the elastic scattering (ES) 
 of electrons.  The CC reaction is sensitive
 exclusively to $\nu_e$'s, while the ES reaction also has 
 a small sensitivity to $\nu_{\mu}$'s and $\nu_{\tau}$'s.  
 The flux of $\nu_e$'s from $^8$B decay measured by 
 the CC reaction rate is  $\phi^{\rm CC}(\nu_e) = 1.75 \pm 0.07~({\rm stat.})^{+0.1
2}_{-0.11}~({\rm sys.}) \pm 0.05~({\rm theor.}) \times 10^6~{\rm cm}^{-2} {\rm s
}^{-1}$. Assuming no flavor transformation,  the flux inferred from the ES
 reaction rate is  $\phi^{\rm ES}(\nu_x)=2.39\pm 0.34~({\rm stat.}) ^{+0.16}_{-0.14}~({\rm sys.}) \times 10^6~{\rm cm}^{-2} {\rm s}^{-1}$.
 Comparison of $\phi^{\rm CC}(\nu_e)$ to the Super-Kamiokande Collaboration's
 precision  value of $\phi^{\rm ES}(\nu_x)$  yields a $3.3\sigma$ difference,
assuming the systematic uncertainties are normally distributed,
 providing evidence that there is a non-electron flavor active
 neutrino component in the solar flux.  The total flux of active  $^8$B 
neutrinos is thus determined to be
 $5.44\pm 0.99 \times 10^6~{\rm cm}^{-2} {\rm s}^{-1}$,
in close agreement with the predictions of solar models.
\end{abstract}


]



Solar neutrino experiments over the past 30 years
\cite{cl,kam,sage,gallex,sk,gno} have 
measured fewer neutrinos than are 
predicted by models of the Sun \cite{BPB,TC}. 
One explanation for the deficit is the transformation of the
Sun's electron-type neutrinos into other active flavors.  The
Sudbury Neutrino Observatory (SNO) 
measures the $^8$B solar neutrinos through the reactions:
 \begin{center}
  \begin{tabular}{ll}
     $\nu_e + d \rightarrow p + p + e^-$\hspace{0.5in} & (CC)\\
     $ \nu_x + d \rightarrow p + n + \nu_x$ & (NC)\\
     $ \nu_x + e^- \rightarrow \nu_x + e^-$  & (ES)\\        
  \end{tabular}
 \end{center}
The charged current reaction (CC) is sensitive exclusively to electron-type
neutrinos, while the neutral current (NC) is sensitive to all active
neutrino flavors ($x=e, \mu, \tau$).  The elastic scattering (ES) reaction
is sensitive to all flavors as well, but with reduced sensitivity to
$\nu_{\mu}$ and $\nu_{\tau}$.  By itself, the ES reaction  cannot provide a 
measure of the total $^8$B flux or its flavor content.
Comparison of the $^8$B flux deduced from the ES reaction assuming
no neutrino oscillations ($\phi^{\rm ES} (\nu_x)$), 
to that measured by the CC reaction ($\phi^{\rm CC}(\nu_e)$) can provide clear 
evidence of flavor transformation without reference to solar model flux 
calculations.
If neutrinos from the Sun change into other active flavors, then
$\phi^{\rm CC}(\nu_e) < \phi^{\rm ES}(\nu_x)$.

This Letter presents the first results from SNO on the ES and CC reactions. 
SNO's measurement of $\phi^{\rm ES}(\nu_x)$ is 
consistent
with previous measurements described in  Ref~\cite{sk}.  The measurement of 
$\phi^{\rm CC}({\nu_e})$, however, is significantly smaller 
and is therefore inconsistent with the null hypothesis that all observed
solar neutrinos are $\nu_e$.  A measurement using the NC reaction, which 
has equal sensitivity to all neutrino flavors, will be reported in a future 
publication.

SNO\cite{NIM} is an imaging water \v{C}erenkov detector
located at a depth of 6010~m of water equivalent in the INCO,
Ltd.\ Creighton mine near Sudbury, Ontario.  It features 
1000 metric tons of ultra-pure D$_2$O contained in a 12~m
diameter spherical acrylic vessel.  This sphere is surrounded by
a shield of ultra-pure H$_2$O contained in a 34~m high
barrel-shaped cavity of maximum diameter 22~m.  A stainless steel 
structure 17.8~m in diameter supports 9456 20-cm photomultiplier
tubes (PMTs) with light concentrators.  Approximately 55\% of the
light produced within 7~m of the center of the detector 
will strike a PMT.

The data reported here were recorded between Nov. 2, 1999 and Jan.
15, 2001 and correspond to a live time of 240.95 days.
Events are defined by a multiplicity trigger of 18 or more PMTs exceeding
a threshold of $\sim{\rm 0.25}$ photo-electrons within a time window
of 93 ns.  The trigger reaches 100\% efficiency at 23 PMTs.
The total instantaneous trigger rate is 15-18 Hz, of which 6-8 Hz is
the data trigger.  For every event trigger, the time and charge responses 
of each participating PMT are recorded.

The data were partitioned into two sets, with approximately 70\%
used to establish the data analysis procedures and 30\% reserved for a blind
test of statistical bias in the analysis.  The analysis procedures were
frozen before the blind data set was analyzed, and no statistically significant
differences in the data sets were found.  We present here the analysis of the
combined data sets.  

Calibration of the PMT time and charge  pedestals, slopes,
offsets, charge vs. time dependencies, and second order rate dependencies are
performed using electronic pulsers and pulsed light sources.
Optical calibration is obtained using a
diffuse source of pulsed laser light at 337, 365, 386, 420, 500 and
620 nm.   The absolute energy scale and uncertainties are established with a 
triggered $^{16}$N source (predominantly 6.13-MeV $\gamma$'s) deployed 
over two planar grids within the D$_2$O and a linear grid in the H$_2$O.  The 
resulting Monte Carlo predictions of detector response are tested using a 
$^{252}$Cf neutron source, which provides an extended distribution of 6.25-MeV 
$\gamma$ rays from neutron capture, and a 
$^3{\rm H}(p,\gamma)^4{\rm He}$ \cite{poon} source providing
19.8-MeV $\gamma$ rays.  The volume-weighted mean response  
is approximately nine PMT hits per MeV of electron energy.

Table~\ref{data_flow} details the steps in data reduction.
The first of these is the elimination of
instrumental backgrounds. Electrical pickup may produce false PMT
hits, while electrical discharges in the PMTs or insulating detector 
materials produce light.  These 
backgrounds have characteristics very different from 
\v{C}erenkov light, and are eliminated using cuts based only
on the PMT positions, the PMT time and charge data, event-to-event
time correlations, and veto PMTs.  This step in the data reduction is
verified by comparing results from two independent background
rejection analyses. 

\begin{table}
\caption{\label{data_flow}  Data reduction steps.}
\begin{tabular}{lr}
Analysis step & Number of events \\
\tableline
Total event triggers      &    355~320~964  \\
Neutrino data triggers    &    143~756~178     \\
$N_{\text{hit}}\geq$30    &    6~372~899 \\
Instrumental background cuts  &  1~842~491     \\
Muon followers     &  1~809~979 \\
High level cuts\tablenote{Reconstruction figures of merit, prompt
light, and  $\langle\theta_{ij}\rangle$.}        &  923~717 \\
Fiducial volume cut         & 17~884 \\
Threshold cut			& 1 169  \\
\tableline 
Total events            &  1 169 \\
\end{tabular}
\end{table}
For events passing the first stage, the calibrated times and positions of
the hit PMTs are used to reconstruct the vertex position and the direction of 
the particle.  The 
reconstruction accuracy and resolution are measured using 
Compton electrons from the 
$^{16}$N source, and the energy and source variation of reconstruction
are checked with a $^8$Li $\beta$ source. 
Angular resolution is measured using Compton electrons produced more
than 150~cm from the $^{16}$N source.  At these energies, the 
vertex resolution is 16~cm and the angular resolution is 
26.7 degrees.

An effective kinetic energy, $T_{\rm eff}$, is assigned to each event 
passing the reconstruction stage.  $T_{\rm eff}$ is calculated using prompt 
(unscattered) \v{C}erenkov photons and the position and direction of
the event.   The derived energy response of the detector can be 
characterized by a Gaussian:
\[
R(E_{\text{eff}},E_{e}) = \frac{1}{ \sqrt{2\pi} \sigma_{E}(E_{e})} \exp[ - \frac{1}{2}(
 \frac{E_{\text{eff}} - E_{e}}{\sigma_{E}(E_{e})} ) ^{2} ]
\]
\noindent where $E_e$ is the total electron energy, 
$E_{\text{eff}} = T_{\rm eff} + m_e$, and 
$\sigma_{E}(E_{e}) = (-0.4620 + 0.5470 \sqrt{E_{e}} + 0.008722 E_{e})$~MeV
is the energy resolution.  The uncertainty on the energy scale is found to be 
$\pm 1.4$\%, which results in a flux uncertainty nearly 4 times larger.  For 
validation, a second energy estimator counts all PMTs hit in each event, 
$N_{\text{hit}}$, without position and direction corrections.

Further instrumental background rejection is obtained using 
reconstruction figures of merit, PMT time residuals, and the
average angle between hit PMTs ($\langle\theta_{ij}\rangle$), measured
from the reconstructed vertex.  These cuts test 
the hypothesis that each event has the characteristics of single electron
\v{C}erenkov light.  The effects of these and the rest of the 
instrumental background removal cuts on neutrino signals are
quantified using the $^8$Li and $^{16}$N sources deployed
throughout the detector.  The volume-weighted neutrino signal loss 
is measured to be $1.4^{+0.7}_{-0.6}$\% and the residual instrumental 
contamination for the data set within the D$_2$O is $< 0.2$\%.  Lastly, cosmic 
ray induced neutrons and spallation products are removed using a 20~s
coincidence window with the parent muon.

\begin{figure}[h]
\begin{center}
\epsfig{figure=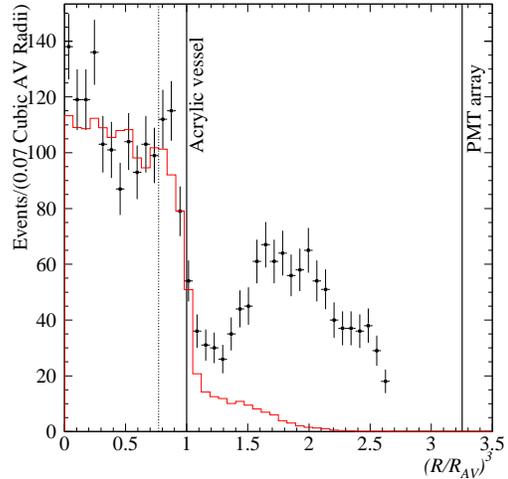,angle=0,height=2.8in}
\end{center}
\caption{\label{fig_1} Distribution of event candidates with
$T_{\rm eff}$$\geq$6.75 MeV as a function of the volume weighted
radial variable 
(R/R$_{\text AV})^3$.  The Monte Carlo simulation of the signals,
weighted by the results from the signal extraction, is shown as a
histogram. The dotted line indicates the fiducial volume cut used in
this analysis.}
\end{figure}

Figure~\ref{fig_1} shows the radial distribution of all remaining events 
above a threshold of $T_{\rm eff}$$\geq$6.75 MeV.  The 
distribution is expressed as a function of the volume-weighted radial 
variable $(R/R_{\rm AV})^3$, where
$R_{\rm AV}=6.00$~m is the radius of the acrylic vessel.
Above this energy threshold, there are contributions from CC events in the 
D$_2$O, ES events in the D$_2$O and H$_2$O, a residual tail
of neutron capture events, and high energy $\gamma$ rays from 
radioactivity in the outer detector.  The data show a clear signal within the 
D$_2$O volume.  For $(R/R_{\rm AV})^3 > 1.0 $ the distribution rises into 
the H$_2$O region until it is cut off by the acceptance of the PMT light 
collectors at $R \sim 7.0$~m.  A fiducial volume cut is applied at $R = 5.50$~m
to reduce backgrounds from regions exterior to the D$_2$O, and to
minimize systematic uncertainties associated with optics and
reconstruction near the acrylic vessel.

Possible backgrounds from radioactivity in the D$_2$O and H$_2$O are
measured by regular low level radio-assays of U and Th decay chain products
in these regions.  The \v{C}erenkov light character of D$_2$O and
H$_2$O radioactivity backgrounds is used {\it in situ} to monitor
backgrounds between radio-assays.  Low energy radioactivity
backgrounds are removed by the high threshold imposed,
as are most neutron capture events.  Monte Carlo calculations predict that 
the H$_2$O shield effectively reduces contributions of low energy 
($<$ 4~MeV) $\gamma$ rays from the PMT array, and these predictions are 
verified by deploying an encapsulated Th source in the vicinity of the
PMT support sphere.  High energy $\gamma$ rays from the cavity are
also attenuated by the H$_2$O shield. A limit on their leakage into 
the fiducial volume is estimated by deploying the $^{16}$N source near
the edge of the detector's active volume.  The total contribution from all 
radioactivity in the detector is found to be $<$0.2\%  for low 
energy backgrounds and $<$0.8\% for high energy backgrounds.

\begin{figure}[h]
\begin{center}
\epsfig{figure=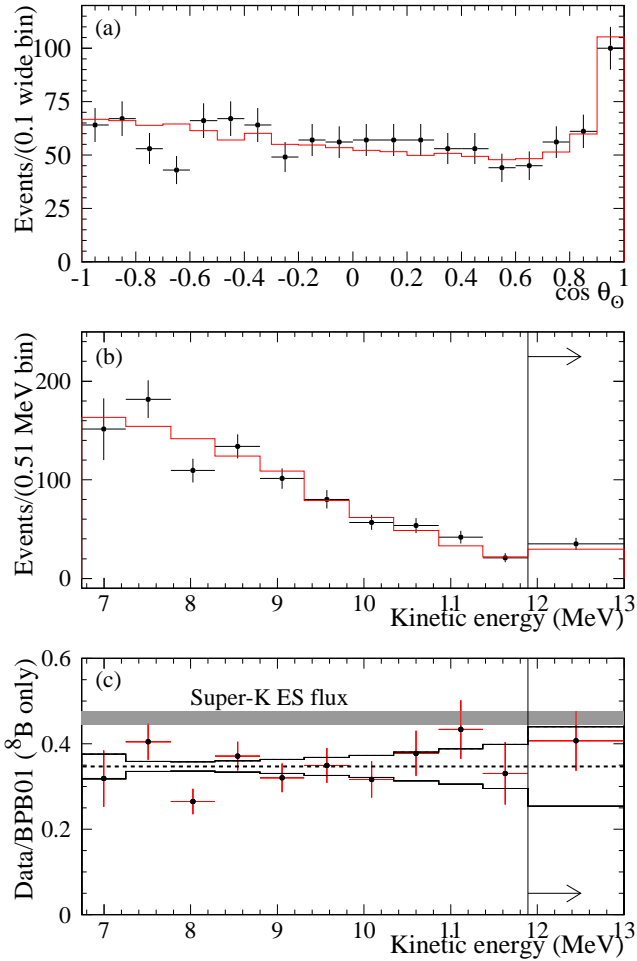,height=5.0in}
\end{center}
\caption{\label{fig_2} Distributions of (a)
$\cos\theta_\odot$, and  (b) extracted kinetic energy spectrum for CC events 
with $R\leq$5.50~m and $T_{\rm eff}$$\geq$6.75 MeV.  The Monte Carlo 
simulations for an
undistorted $^8$B spectrum are shown as histograms. The ratio of the
data to the expected kinetic energy distribution with correlated
systematic errors is shown in (c). The uncertainties in the $^8$B 
spectrum{\protect \cite{ortiz}} have not been included.}
\end{figure}

The final data set contains 1169 events after the fiducial volume and
kinetic energy threshold cuts.  Figure~\ref{fig_2} (a) displays the
distribution of $\cos\theta_\odot$,  the angle
between the reconstructed direction of the event and the instantaneous
direction from the Sun to the Earth.  The forward peak in this
distribution arises from the kinematics of the ES reaction, while
CC electrons are expected to have a distribution which is 
$(1-0.340\cos\theta_\odot)$~\cite{vogel}, before accounting for detector 
response.  

The data are resolved into contributions from CC, ES, and neutron
events above threshold using probability density functions
(pdfs) in $T_{\text{eff}}$, $\cos\theta_\odot$, and $(R/R_{\rm AV})^3$,
generated from Monte Carlo simulations assuming no flavor transformation
and the shape of the standard $^8$B spectrum~\cite{ortiz} ({\it hep}
neutrinos are not included in the fit).   The
extended maximum likelihood method used in the signal extraction yields
975.4$\pm$39.7 CC 
events, 106.1$\pm$15.2 ES events, and
87.5$\pm$24.7  neutron events for
the fiducial volume and the threshold chosen, where the uncertainties given are
statistical only.  The dominant
sources of systematic uncertainty in this signal extraction are the energy
scale uncertainty and reconstruction accuracy, as shown in
Table~\ref{errors}.  The CC and ES signal
decomposition gives consistent results when used with the
$N_{\text{hit}}$ energy estimator, as well as with different choices of the
analysis threshold and the fiducial volume up to 6.20~m with
backgrounds characterized by pdfs.
\begin{table}
\caption{\label{errors} Systematic error on fluxes.}
\begin{tabular}{ldd}
Error source  & CC  error  & ES  error\\
		& (percent) & (per cent)\\
\tableline
Energy scale		& -5.2, +6.1  & -3.5 ,+5.4 \\
Energy resolution	& $\pm$0.5	 & $\pm$0.3	  \\
Energy scale non-linearity           & $\pm$0.5       & $\pm$0.4 \\
Vertex accuracy		& $\pm$3.1	 & $\pm$3.3	  \\
Vertex resolution	& $\pm$0.7       & $\pm$0.4	  \\
Angular resolution	& $\pm$0.5	 & $\pm$2.2	  \\
High energy $\gamma$'s  & -0.8, +0.0     & -1.9, +0.0  \\
Low energy background   &  -0.2, +0.0     & -0.2, +0.0 \\
Instrumental background &  -0.2, +0.0    & -0.6, +0.0 \\
Trigger efficiency      &  0.0           & 0.0 \\
Live time		& $\pm$0.1	 & $\pm$0.1	  \\
Cut acceptance		& -0.6, +0.7     & -0.6, +0.7  \\
Earth orbit eccentricity & $\pm$0.1      & $\pm$0.1 \\
$^{17}$O, $^{18}$O      &  0.0           &  0.0 \\
\tableline
Experimental uncertainty  & -6.2, +7.0     & -5.7, +6.8  \\
\tableline
Cross section		& 3.0	& 0.5	  \\
\tableline
Solar Model             & -16, +20  & -16, +20  \\
\end{tabular}
\end{table}

The CC spectrum can be extracted from the data by removing the constraint
on the shape of the CC pdf and repeating the signal extraction.

Figure~\ref{fig_2} (b) shows the kinetic energy spectrum with statistical
error bars, with the $^8$B 
spectrum of Ortiz~{\em et al.}~\cite{ortiz} scaled to the data. 
The ratio of the data to the prediction~\cite{BPB}
is shown in Figure~\ref{fig_2} (c).  The bands
represent the 1$\sigma$ uncertainties derived from the most significant
energy-dependent systematic errors.  
There is no evidence for a deviation of the
spectral shape from the predicted shape under the non-oscillation hypothesis.

Normalized to the integrated rates above the kinetic energy threshold
of $T_{\text{eff}} = 6.75$~MeV, the  measured $^8$B neutrino fluxes 
assuming the standard spectrum shape~\cite{ortiz} are:
\begin{eqnarray*}
\phi_{\text{SNO}}^{\text{CC}}(\nu_e) & = & 1.75 \pm 0.07~({\rm stat.})
      ^{+0.12}_{-0.11}~({\rm sys.}) \pm 0.05~({\rm theor.}) \\
                              &   & \times 10^6~{\rm cm}^{-2} {\rm s}^{-1} \\
\phi_{\text{SNO}}^{\text{ES}}(\nu_x) & = & 2.39 \pm 0.34 ({\rm stat.})
      ^{+0.16}_{-0.14}~({\rm sys.}) \times 10^6~{\rm cm}^{-2} {\rm s}^{-1}
\end{eqnarray*}
where the theoretical uncertainty is the CC cross section 
uncertainty\cite{xsect}.  Radiative corrections have not been applied
to the CC cross section, but they are expected to decrease the measured 
$\phi^{\rm CC}(\nu_e)$ flux\cite{rad_corr} by up to a few percent.
The difference between the $^8$B flux
deduced from the ES rate and that deduced from the CC rate in SNO is
$0.64\pm 0.40 \times 10^{6}~{\rm cm}^{-2} {\rm s}^{-1}$, or
1.6$\sigma$. SNO's ES rate measurement is consistent with the 
precision measurement by the Super-Kamiokande Collaboration of the $^8$B flux 
using the same ES reaction~\cite{sk}:
\begin{displaymath}
\phi_{\text{SK}}^{\text{ES}}(\nu_x) = 2.32 \pm 0.03~({\rm stat.})
^{+0.08}_{-0.07}~({\rm sys.}) \times 10^6~{\rm cm}^{-2} {\rm s}^{-1}.
\end{displaymath}
The difference between the flux $\phi^{\text{\rm ES}}(\nu_x)$ 
measured by Super-Kamiokande 
via the ES reaction and the $\phi^{\text CC}(\nu_e)$ flux measured by SNO via
the CC reaction is $0.57\pm 0.17 \times 10^{6}$ cm$^{-2}{\rm s}^{-1}$, or 
3.3$\sigma$~\cite{hep}, assuming that the systematic errors are normally
distributed. The probability that a downward fluctuation of the Super-Kamiokande
result would produce a SNO result $\ge 3.3 \sigma$ is 0.04\%.
For reference, the ratio of the SNO CC $^8$B flux to that of the BPB01
solar model\cite{BPB} is 0.347$\pm$0.029, where all uncertainties 
are added in quadrature.

If oscillation solely to a sterile neutrino is occurring,
the SNO CC-derived $^8$B flux above a threshold of 6.75 MeV will be 
consistent with the integrated Super-Kamiokande ES-derived 
$^8$B flux above a threshold of 8.5 MeV\cite{lisi1}. Adjusting the ES 
threshold\cite{sk} this derived flux difference is $0.53\pm 0.17 \times 10^{6}$ cm$^{-2}{\rm s}^{-1}$, or 
3.1$\sigma$.  The probability of a downward fluctuation $\ge 3.1 \sigma$ is 
0.13\%.  These data are therefore evidence of a non-electron
active flavor component in the solar neutrino flux.  These data are also 
inconsistent with the ``Just-So$^2$'' 
parameters for neutrino oscillation~\cite{BKS}.  

\begin{figure}[h]
\begin{center}
\epsfig{figure=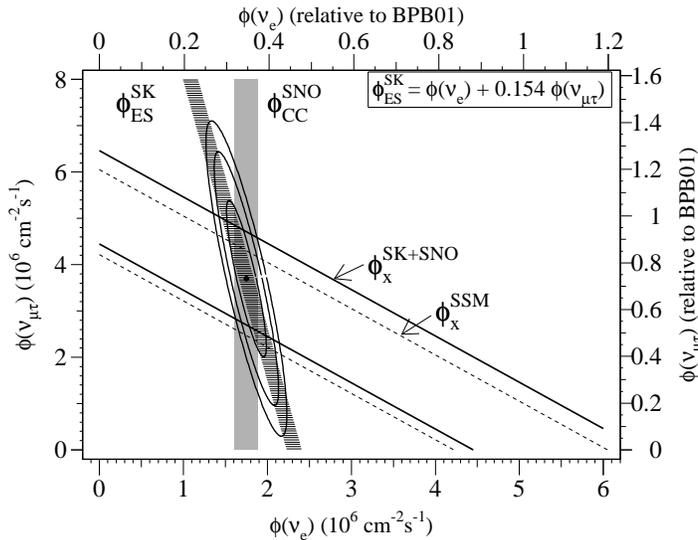,angle=0,height=2.8in}
\end{center}
\caption{\label{fig_3} Flux of $^8$B solar neutrinos
which are $\mu$ or $\tau$ flavor vs. the flux of electron
neutrinos as deduced from the SNO and Super-Kamiokande data. 
The diagonal bands show the total $^8$B flux $\phi(\nu_x)$ as predicted by 
BPB01 (dashed lines) and that derived from the SNO and Super-Kamiokande 
measurements (solid lines). The intercepts of these bands with the axes 
represent the $\pm 1 \sigma$ errors.}
\end{figure}

Figure~\ref{fig_3} displays the inferred flux of non-electron flavor active 
neutrinos ($\phi(\nu_{\mu \tau})$)
against the flux of electron neutrinos. The two data bands represent the one 
standard deviation measurements of the SNO CC rate and the Super-Kamiokande 
ES rate. The error ellipses represent the 68\%, 95\%, and 99\% joint
probability contours for $\phi(\nu_e)$ and 
$\phi({\nu_{\mu \tau}})$.
The best fit to $\phi({\nu_{\mu \tau}})$ is:
\begin{displaymath}
\phi(\nu_{\mu \tau}) =  3.69\pm 1.13 \times 10^6~{\rm cm}^{-2} 
{\rm s}^{-1}.
\end{displaymath}

The total flux of active $^8$B neutrinos is determined to be: 
\begin{displaymath}
\phi(\nu_x) =  5.44\pm 0.99 \times 10^6~{\rm cm}^{-2}
{\rm s}^{-1}. 
\end{displaymath}
This result is displayed as a diagonal band in Fig.~\ref{fig_3}, and
is in excellent agreement with predictions of standard solar
models~\cite{BPB,TC}.

Assuming that the oscillation of massive neutrinos explains both the evidence 
for electron neutrino flavor change presented here and the atmospheric 
neutrino data of the Super-Kamiokande collaboration~\cite{SKatm}, two 
separate splittings of the squares of the neutrino mass eigenvalues
are indicated: $< 10^{-3} {\text{eV}}^2$ for the 
solar sector~\cite{CHOOZ,BKS} and $\simeq 3.5 \times 10^{-3} {\text{eV}}^2$
for atmospheric neutrinos.  These 
results, together with the beta spectrum of tritium~\cite{Mainz}, 
limit the sum of mass eigenvalues of active neutrinos to be between 
0.05 and 8.4 eV, corresponding to a constraint of $0.001 < \Omega_\nu <0.18$ 
for the  contribution to the critical density of the 
Universe~\cite{cosmo,Weiler}.

In summary, the results presented here are the first direct indication 
of a non-electron flavor component in the solar neutrino flux, and enable   
the first determination of the total flux of $^8$B neutrinos generated by the 
Sun.

This research was supported by the Natural
Sciences and Engineering Research Council of Canada, Industry Canada, National
Research Council of Canada, Northern Ontario Heritage Fund Corporation
and the Province of Ontario,  the United States Department of
Energy, and in the United Kingdom by the Science and Engineering
Research Council and the Particle Physics and Astronomy Research
Council.  Further support was provided by INCO, Ltd., Atomic Energy of
Canada Limited (AECL), Agra-Monenco, Canatom, Canadian
Microelectronics Corporation, AT\&T Microelectronics, Northern Telecom and 
British Nuclear Fuels, Ltd.   The heavy water
was loaned by AECL with the cooperation of Ontario Power Generation.


\begin{references}
\bibitem{hep} Given the limit set for the {\it hep} flux by
Ref.~\cite{sk}, the effects of the {\it hep} contribution may
increase this difference by a few percent.
\end{references}
\end{document}